\documentstyle[12pt,epsf]{article}

\ifx\epsffile\undefined
\newlength{\epsfysize}
\def\epsffile#1#2#3#4]#5{}
\fi

\newcommand{\be}{\begin{equation}}
\newcommand{\ee}{\end{equation}}
\newcommand{\bear}{\begin{eqnarray}}
\newcommand{\eear}{\end{eqnarray}}
\newcommand{\gsim}{\lower.7ex\hbox{$\;\stackrel{\textstyle>}{\sim}\;$}}
\newcommand{\lsim}{\lower.7ex\hbox{$\;\stackrel{\textstyle<}{\sim}\;$}}

\newcommand{\met}{\rlap{\,/}E_T}

\newcommand{\goldstino}{\tilde{G}}
\def\npb#1 #2 #3 #4 {Nucl.~Phys. {\bf B#1}, #2 (#3)#4 }
\def\plb#1 #2 #3 #4 {Phys.~Lett. {\bf B#1}, #2 (#3)#4 }
\def\prd#1 #2 #3 #4 {Phys.~Rev.  {\bf D#1}, #2 (#3)#4 }
\def\prl#1 #2 #3 #4 {Phys.~Rev.~Lett. {\bf #1}, #2 (#3)#4 }
\def\pr#1  #2 #3 #4 {Phys.~Rept. {\bf #1}, #2 (#3)#4 }
\def\mpl#1 #2 #3 #4 {Mod.~Phys.~Lett. {\bf A#1}, #2 (#3)#4 }
\def\zpc#1 #2 #3 #4 {Z.~Phys. {\bf C#1}, #2 (#3)#4 }

\begin{document}

\baselineskip=24pt

\thispagestyle{empty}
\begin{titlepage}
\title {\vspace{-2.0cm} 
\hfill {\normalsize FERMILAB--PUB--99/140--T}\\
\hfill {\normalsize SU-ITP-99-35}\\
\hfill {\normalsize August 26, 1999}\\
\vspace{3.0cm} 
Higgs and $Z$-boson Signatures of Supersymmetry}

\vspace{3.5cm}

\author{\\
{\sc Konstantin T. Matchev}\\
{\small Theoretical Physics Department}\\
{\small Fermi National Accelerator Laboratory}\\
{\small Batavia, IL 60510}\\
\\
{\sc Scott Thomas} \\ 
{\small Physics Department} \\
{\small Stanford University} \\
{\small Stanford, CA 94305}\\
}

\date{}
\maketitle
\thispagestyle{empty}

\begin{abstract}
\noindent
In supersymmetric theories of nature the Higgsino fermionic
superpartner of the Higgs boson can arise as the lightest
standard model superpartner depending on the couplings
between the Higgs and supersymmetry breaking sectors.
In this letter the production and decay of Higgsino pairs
to the Goldstone fermion of supersymmetry breaking and the
Higgs boson, $h$, or gauge bosons, $Z$ or $\gamma$ are considered.   
Relatively clean di-boson final states, $hh$, $h \gamma$, $hZ$,
$Z \gamma$, or $ZZ$, with a large amount of missing energy result.
The latter channels provide novel discovery modes for supersymmetry
at high energy colliders since events with $Z$ bosons are generally
rejected in supersymmetry searches. In addition, final states with
real Higgs bosons can potentially provide efficient channels
to discover and study a Higgs signal at the Fermilab Tevatron Run II.
\end{abstract}

\vspace{1.8cm}

\end{titlepage}

\setcounter{page}{1} 

\section{Introduction}

Supersymmetry (SUSY) provides perhaps the best motivated extension 
of the Standard Model. 
Spontaneous SUSY breaking leads naturally to radiative
electroweak symmetry breaking with masses of order the electroweak scale
for the superpartners of the Standard Model (SM) particles. 
If the messenger interactions which couple the SM
superpartners to the SUSY breaking
sector are stronger than gravity, the lightest supersymmetric 
particle (LSP) is the Goldstone fermion of supersymmetry
breaking, the Goldstino $\tilde G$.
The next to lightest supersymmetric particle
(NLSP) is generally the lightest SM superpartner.
If the intrinsic scale of supersymmetry breaking 
is below $\sim 10^3$ TeV the NLSP can decay to its 
SM partner and the Goldstino on 
laboratory length scales \cite{goldstino}.
This has an important impact on experimental SUSY signatures
at high energy colliders.
Since superpartners are generally produced in pairs, 
these decays give rise to final states with 
two hard partons and missing energy ($\met$)
carried by the Goldstino 
pair, and with possibly other partons in the final state from cascade
decays to the NLSP \cite{goldstino,supp,review}.

The identity of the NLSP determines the type of final 
states which arise from decay to the Goldstino \cite{review}.
A neutralino NLSP, 
$\tilde \chi_1^0$, which is gaugino-like, can decay by 
$\tilde \chi_1^0 \rightarrow \gamma  \tilde G$, leading to final 
states with $\gamma \gamma  \met$. A slepton NLSP, 
$\tilde \ell$, can decay by $\tilde \ell \rightarrow \ell  \tilde G$, 
giving $\ell\ell  \met$ final states. 
In this letter we consider in detail the possibility
of a fermionic Higgsino-like neutralino NLSP.
Because it is the superpartner of the Higgs boson, $h$,  
a Higgsino NLSP can decay by 
$\tilde \chi_1^0 \rightarrow h  \tilde G$. 
In addition, since the longitudinal component of the $Z$ boson 
mixes with the Goldstone mode of the Higgs field, 
$\tilde \chi_1^0 \rightarrow Z  \tilde G $ can also result.
Because of a strong phase space suppression of 
the $h$ and $Z$ final states near threshold, decay to a photon
can also be important for 
Higgsinos not too much heavier than the $Z$ boson.
Pair production of Higgsinos which decay to Goldstinos 
can then give rise to the di-boson final states
$(hh,h \gamma, hZ, Z \gamma, ZZ)  \met$ \cite{talks}.

Di-boson signatures which include Higgs and 
$Z$ bosons and $\met$ are quite novel discovery modes
for supersymmetry in the mass range accessible to the current
generation of high energy collider experiments. 
In conventional SUSY signatures, in which the lightest neutralino, 
$\tilde \chi_1^0$, is assumed to escape the detector without
decay to the Goldstino, the mass splittings between supersymmetric
particles required in order for $h$ or $Z$ to arise in a cascade
decay, typically imply the superpartners are too heavy to be
produced in sufficient numbers at present colliders. 
For this reason events with reconstructed $Z$ bosons are in
fact generally rejected in present SUSY searches. 
However, since the Goldstino is essentially massless, sufficient
phase space is available for the $h \goldstino$ and 
$Z \goldstino$ modes for a Higgsino somewhat heavier than $h$ or $Z$. 
And this mass range will be accessible at the upcoming Run II
at the Fermilab Tevatron. The Higgs final states also present the
exciting possibility of discovering and studying the 
Higgs boson in association with supersymmetry. 

If supersymmetry is broken at a low scale, as required for
the di-boson sigatures discussed here, it is very likely that 
the SM gauge interactions play some role in coupling the SUSY
breaking sector to the SM superpartners \cite{supp}.
However, such gauge-mediated SUSY breaking requires
additional interactions between the Higgs and SUSY 
breaking sectors in order to break certain Higgs sector global 
symmetries and obtain acceptable electroweak symmetry 
breaking \cite{review}. These interactions can modify the Higgsino
mass from minimal expectations, and allow for a Higgsino NLSP.
So searches for di-boson signatures of a Higgsino NLSP within 
theories of low scale gauge-mediated SUSY breaking 
are very well motivated as possible indirect probes for the existence of 
these additional couplings. 


\section{Higgsino decays and production}
The Higgsinos $\tilde H_u$ and $\tilde H_d$ are fermionic
superpartners of the Higgs boson fields $H_u$ and $H_d$.
The neutral Higgsinos mix with the gaugino superpartners
of the $\gamma$ and $Z$ gauge bosons, while the charged 
Higgsino mixes with the gaugino superpartner of the $W$
gauge boson. In the limit relevant here,
in which the gauginos are heavier than the Higgsinos,
the two lightest neutralinos and lightest chargino,
$\tilde \chi_1^0$, $\tilde \chi_2^0$, $\tilde \chi_1^{\pm}$, 
are predominantly Higgsino and approximately degenerate. 
The splitting between these states is on the order of 10-15 
GeV for masses in the range 120-250 GeV discussed below. 
If the $U(1)_Y$ and $SU(2)_L$ gaugino mass parameters, 
$M_1$ and $M_2$, have the same sign, ${\rm sgn}(M_1 M_2)=+$
then $\tilde \chi_1^0$ is the NLSP. 
For ${\rm sgn}(M_1 M_2) = -$ it is however possible 
in certain regions of parameter space that 
$\tilde \chi_1^{\pm}$ is the NLSP.
In this letter only a $\tilde \chi_1^0$ NLSP,
which leads to the interesting
di-boson signatures, will be considered. 

The branching ratios 
${\rm Br}(\tilde \chi^0_1 \rightarrow \tilde G + (\gamma,h,Z))$
are determined by the Higgsino and gaugino content 
of $\tilde \chi_1^0$ \cite{review,BR-NLSP}. 
\begin{figure}[t!]
\epsfysize=3.5in
\epsffile[-40 220 170 570]{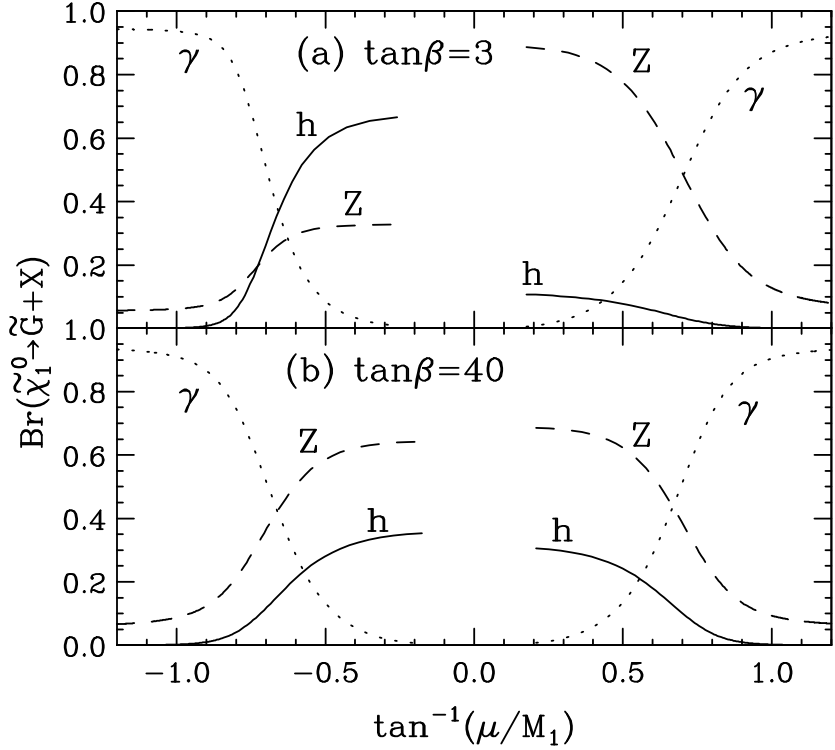}
\begin{center}
\parbox{5.5in}{   
\caption[]{Branching ratios of the lightest neutralino
${\rm Br}(\tilde \chi^0_1 \rightarrow \tilde G + \gamma,h,Z)$ 
as a function of the neutralino mixing angle $\tan^{-1}(\mu/M_1)$,
for a fixed mass $M_{\tilde \chi^0_1}=160$ GeV 
and $m_h=105$ GeV for (a) $\tan\beta=3$ and (b) $\tan\beta=40$. 
\label{br}}}
\end{center}
\end{figure}
This is illustrated in Fig.~\ref{br} as a function of 
the neutralino mixing angle $\tan^{-1}(\mu/M_1)$ for fixed 
$\tilde \chi_1^0$ mass, where $\mu$ is the Higgsino mass parameter, 
and $\tan \beta = v_u / v_d$ is the ratio of Higgs expectation values.  
For definiteness the Higgs decoupling limit
in which decays to the heavy scalar and pseudoscalar Higgs bosons, 
$H$ and $A$, are kinematically blocked is employed throughout. 
For gaugino-like $\tilde \chi_1^0$ the $\gamma$ mode dominates, 
but for Higgsino-like $\tilde \chi_1^0$ the $h$ and $Z$ modes
become important. The dependence on ${\rm sgn}(\mu)$ and $\tan \beta$ 
apparent in Fig.~\ref{br} can be understood
in terms of the $\tilde \chi_1^0$ quantum numbers and couplings
and will be presented elsewhere. 

The branching ratios also depend on the $\tilde \chi_1^0$ 
mass through the phase space available to the $h$ and $Z$ modes
which suffer a $\beta^4$ velocity suppression near 
threshold \cite{review,BR-NLSP}.
So even a Higgsino-like $\tilde \chi_1^0$ decays predominantly
by $\tilde \chi_1^0 \rightarrow \gamma \tilde G$ for masses not 
too far above the $h$ and $Z$ masses. 
\begin{figure}[t!]
\epsfysize=3.5in
\epsffile[-40 220 170 570]{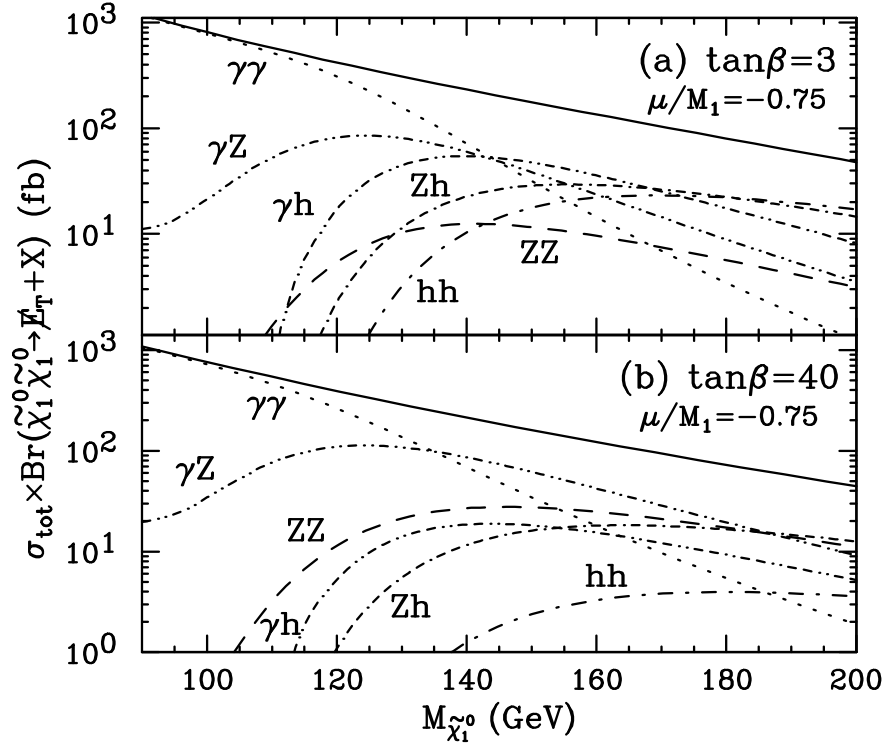}
\begin{center}
\parbox{5.5in}{
\caption[]{Total cross-section 
$\sigma_{tot}(p \bar{p} \rightarrow \tilde{\chi}_i^0 \tilde{\chi}_j^0, 
\tilde{\chi}^{+}_1 \tilde{\chi}^{-}_1)$ in fb for $i,j=1,2$
times the branching ratio into various di-boson final states
as a function of the lightest Higgsino mass $M_{\tilde \chi^0_1}$, 
for a fixed ratio $\mu/M_1 = -3/4$, with
(a) $\tan\beta=3$ or (b) $\tan\beta=40$.
The center of mass energy is 2 TeV and $m_h = 105$ GeV. 
The solid line indicates the total cross section. 
\label{sigbr}}}
\end{center}
\end{figure}
The mass dependence of the branching ratios is illustrated
in Fig.~\ref{sigbr} in which the $p \bar{p}$
signal cross section 
times branching ratio into the di-boson final states
is given as a function of the $\tilde \chi_1^0$ mass for fixed
Higgsino-neutralino mixing.
With $\tilde \chi_1^0$ Higgsino-like the $hh$, $ZZ$, or $hZ$ modes
dominate for very large masses, while the $\gamma \gamma$ mode
dominates for smaller masses. 
However, because of the strong phase space suppression near threshold
there is a transition region which extends over a significant
range of mass between these limits 
in which the mixed final states $\gamma h$ and/or 
$\gamma Z$ (depending on ${\rm sgn(\mu)}$ and $\tan \beta)$
are important. 
These final states are particularly useful for masses in the transition
region since the photon is quite hard.

The total cross section 
$\sigma_{tot}(p \bar{p} \rightarrow \tilde{\chi}_i^0 \tilde{\chi}_j^0, 
\tilde{\chi}^{+}_1 \tilde{\chi}^{-}_1)$ for $i,j=1,2$ in Fig.~\ref{sigbr},
summed over all the Higgsino-like states, is the relevant
signal cross section since these states are approximately
degenerate, and can all be produced at similar rates. 
The heavier states cascade decay to $\tilde \chi_1^0$
through neutral and charged current interactions. 
The partons from these cascade decays are relatively soft 
and probably not particularly useful at the trigger level.


\section{$Z$ boson final states}
The final states with a $Z$ boson can be significant for large
$\tan \beta$, or at small $\tan \beta$ with $\mu > 0$. 
The $Z$ boson can decay invisibly, leptonically, or hadronically, 
$Z \rightarrow \nu \nu, \ell\ell, jj$,
leading to many possible signatures. 
The $ee$ and $\mu \mu$ leptonic decays allow the possibility
of precise reconstruction of the $Z$ invariant mass, but suffer
from small branching ratio,
${\rm Br}(Z \rightarrow ee,\mu \mu) \simeq 6.7 \%$. 
In contrast, the invisible and hadronic decay modes can be useful
because of larger branching ratios, 
${\rm Br}(Z \rightarrow \nu \nu) \simeq 20 \%$, and 
${\rm Br}(Z \rightarrow jj) \simeq 70 \%$.

The $\gamma Z \met$ di-boson mode dominates the total
cross section in the transition region of masses as shown in Fig. 2(b). 
Leptonic decay of the $Z$ provides the cleanest final state, 
$\gamma \ell^+ \ell^- \met$, which is similar to existing 
SM $Z\gamma$ studies without $\met$ \cite{Zgamma_CDF,Zgamma_D0}.
For a Higgsino search, however, an additional large $\met$ cut,
as well as a more stringent photon $E_T$ cut should reduce
the backgrounds to a negligible level.
Our Monte Carlo estimates indicate
that this channel is practically background free, but
is limited by the small leptonic branching ratio of the $Z$ boson. 
The Tevatron Run IIa with $2$ fb$^{-1}$ of integrated luminosity
will have a reach at the 3$\sigma$ discovery level for
$\tilde \chi_1^0$ masses up to 155 GeV for the parameters of Fig. 2(b), 
while the reach in Run IIb with $30$ fb$^{-1}$ should approach 220 GeV. 

Invisible decay of the $Z$ gives rise to the signature $\gamma \met$. 
This channel has been studied in Run I as a probe for anomalous
$\gamma Z$ couplings \cite{photon_D0,diboson_review_D0}.
Backgrounds include $\gamma j$ and $jj$ with one jet faking a photon 
{\it and} in each case the remaining jet energy mismeasured to be below
the minimum pedastool. The largest background in Run I was
from single $W$ production with $W \rightarrow e \nu$ and
the electron misidentified as a photon. 
This background can be substantially reduced by
raising the photon $E_T$ and $\met$ cuts above
50 GeV, beyond the Jacobian peak for 
$W\rightarrow \ell\nu$ \cite{Greg}. 
This also reduces the hadronic background. 
The 3$\sigma$ discovery reach in $\chi_1^0$ mass
should then approach 150 (185) GeV in Run IIa (IIb)
for the parameters of Fig.~\ref{sigbr}(b).

Hadronic decay of the $Z$ in the $\gamma Z\met$ mode gives
rise to the signature $\gamma jj \met$. 
Backgrounds are similar to those of the
$\gamma \met$ channel.  The $\gamma jj \met$ 
channel has been studied in Run I in order to place limits
on squark and gluino masses in very specific supersymmetric
models \cite{photon_jets_D0}.
Further background suppressions not included in 
the Run I study are possible with acoplanarity, sphericity and invariant
dijet mass cuts to reconstruct the $Z$ boson, and a lepton veto.
In any case, the total background is expected to be smaller than 
for the $\gamma\met$ channel, due to the presence of two additional
hard partons. 
Given the significant $Z$ hadronic branching ratio, the $\gamma jj \met$
channel should provide somewhat better reach than the 
$\gamma \ell^+\ell^- \met$ or $\gamma \met$ channels in Run II. 

The $ZZ \met $ di-boson mode dominates at
larger $\tilde \chi_1^0$ mass as shown in Fig.~2(b). 
Leptonic decay of each $Z$ boson gives rise to the spectacular
signature $\ell^+\ell^-\ell^{\prime +} \ell^{\prime -} \met$,
with the lepton pairs
reconstructing the $Z$ mass (in one choice of pairing for
$\ell=\ell^{\prime}$).
This channel is expected to be essentially background free, 
but suffers from small leptonic branching ratio. 
Because of this Run IIb will not be sensitive to this channel for the 
paramters of Fig.~\ref{sigbr}.
But for $\mu / M_1 = 1/3$ and 
$\tan \beta = 3$ with larger 
${\rm Br}(\tilde \chi_1^0 \rightarrow Z \tilde G)$ 
(c.f. Fig.~\ref{br}), the 3$\sigma$ discovery reach in Run IIb
for the $\tilde \chi_1^0$ mass is 170 GeV. 
At the LHC $\ell^+\ell^-\ell^{\prime +} \ell^{\prime -} \met$ 
would represent the gold plated 
channel for the $ZZ\met$ di-boson mode from Higgsino decay. 

Hadronic decay of one of the $Z$ bosons
gives the signature $\ell^+\ell^-jj \met$. 
An important background in this channel comes from 
$t\bar{t}$ production with $t \rightarrow Wb$
and $W \rightarrow \ell \nu$
with the $\ell^+\ell^-$ pair reconstructing the $Z$ mass, 
and each $b$-jet not identified as a heavy flavor. 
Other backgrounds arise from $ZZ$ and $WZ$ 
in association with jets.
In Run IIb the 3$\sigma$ discovery reach in $\tilde \chi_1^0$ mass
should approach 195 GeV for
$\mu / M_1 = 1/3$ and $\tan \beta =3$. 
Rejecting backgrounds for the other 
decay channels of the $ZZ\met$ di-boson mode presents
more serious challenges. 


\section{Higgs boson final states}
The decay of Higgsinos to real Higgs bosons gives perhaps
the most interesting di-boson final states because of the 
opportunity to study both supersymmetry and the Higgs sector. 
Higgs boson final states are important for small
$\tan \beta$ and $\mu <0$ or for large $\tan \beta$ with 
sufficiently large $\tilde \chi_1^0$ mass, as shown in 
Figs.~\ref{br} and \ref{sigbr}. 

In the transition region of $\tilde \chi_1^0$ mass,
$\gamma h \met$ is the most important di-boson mode. 
With the dominant decay $h \rightarrow bb$ this 
leads to the signature $\gamma bb \met$. 
Backgrounds include 
$Z \gamma j$ and $Z jj$ with $Z \rightarrow bb$
and $bb \gamma j$ and $bbjj$ with one jet misidentified 
as a photon {\it and} in each case the remaining jet energy 
mismeasured
to be below the minimum pedastool.
Based on the work presented here \cite{talks}
it has been estimated \cite{report} 
that with a single $b$-tag the 3$\sigma$ discovery reach in 
$\tilde \chi_1^0$ mass should approach
210 (250) GeV in Run IIa (IIb) 
for the parameters of Fig.~\ref{sigbr}(a).

For larger $\tilde \chi_1^0$ masses the $hZ\met$ and/or $hh\met$ modes 
can become important, as shown in Fig.~\ref{sigbr}. 
The $hh \met$ di-boson final state gives rise to the 
signature $bbbb \met$.
The sizeable QCD and electroweak backgrounds to this final state can 
be significantly reduced by requiring at least 3 tagged 
$b$-jets with large invariant mass for two $b$-jet pairs
\cite{talks}, as verified by Monte Carlo 
simulation \cite{bbbb}. Remaining backgrounds include 
$ZZj$ with each $Z \rightarrow bb$,
$bbjj$ with one jet misidentified as a $b$-jet, and 
$bbbbj$ with in each case the jet energy 
mismeasured to be below the minimum pedastool, and 
$t \bar{t}$ production with $t \rightarrow Wb$ and 
one hadronic decay 
$W \rightarrow jj$ with one jet misidentified as a $b$-jet, 
and one leptonic decay $W \rightarrow \ell \nu$ with 
$\ell$ not identified. 
Accounting for the $t\bar{t}$ background \cite{bbbb},
the 3$\sigma$ discovery reach in $\tilde \chi_1^0$ mass
at Run IIb should approach 240 GeV for the parameters of 
Fig.~\ref{sigbr}(a). 

The $hZ\met$ di-boson mode arising from Higgsino decay 
is similar to direct $hZ$ production.
Invisible decay of the $Z$ gives the signature $bb \met$, 
and would contribute slightly to searches for the SM Higgs
boson in this channel. Leptonic decay of the $Z$ gives the signature
$\ell^+\ell^-bb\met$. Unfortunately, the dominant background 
from $t\bar{t}$ production with 
$t\rightarrow Wb$ and $W \rightarrow \ell \nu$ with 
the $\ell^+\ell^-$ pair reconstructing the $Z$ mass, 
is very similar to the signal. 
Because of this, Run II is not expected to be sensitive to this channel. 
Hadronic decay with $Z \rightarrow bb$ gives the signature $bbbb \met$, 
similar to the $hh\met$ mode. 
However, because of the smaller branching ratio, 
${\rm Br}(Z \rightarrow bb) / {\rm Br}(h \rightarrow bb) \simeq
20\%$, Run II will just marginally 
not be sensitive to the $hZ\met$ mode in this channel
for the parameters of Fig.~\ref{sigbr}. 

The Higgs boson final states of Higgsino decay discussed above
present the possibility of collecting
a relatively clean sample of events which contain real Higgs bosons. 
It is therefore interesting to consider the reach
as a general function of both Higgsino and Higgs masses. 
The total cross section times branching ratio contours for the 
$\gamma bb \met$ and $bbbb \met$ channels as a function 
of the $h$ and $\tilde \chi_1^0$  masses are shown in Fig. 3.
\begin{figure}[t!]
\epsfysize=3.5in
\epsffile[-10 240 200 570]{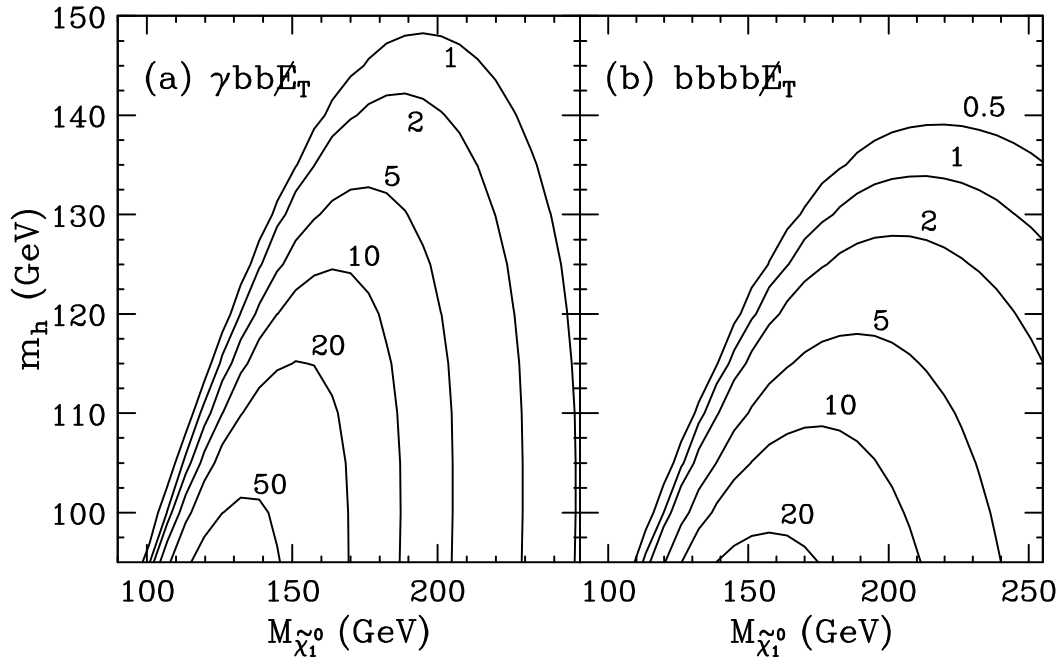}
\begin{center}
\parbox{5.5in}{
\caption[]{ Signal cross-section times branching ratio
contours in fb
for the (a) $\gamma bb\met$ and (b) $bbbb\met$ channels,
as a function of the neutralino mass
$M_{\tilde \chi^0_1}$, and the Higgs mass $m_h$,
for $\tan\beta=3$ and $\mu/M_1=-3/4$.
\label{higgs xsections}}}
\end{center}
\end{figure}
These contours include ${\rm Br}(\chi_1^0 \to (\gamma, h) \tilde G)$
for $\tan\beta=3$ and $\mu/M_1=-3/4$ and SM
values for ${\rm Br}(h \to bb)$.
The Run IIa 3$\sigma$ dicovery reach quoted above for the 
$\gamma bb \met$ channel corresponds to a 
signal times branching ratio cross section of 5 fb.
For the parameters of Fig. 3 this corresponds to a Higgs
mass of up to at least 120 GeV for $\tilde \chi_1^0$ masses in the 
range 135-200 GeV, 
with a maximum reach in Higgs mass of just over 130 GeV. 
This is to be contrasted with the search
for the SM Higgs from direct $Wh$ and $Zh$ production.
These SM channels are background limited, and 
no sensitivity to a Higgs mass  
beyond current limits is expected in Run IIa \cite{higgsreport}.
So the $\gamma bb \met$ channel presents the interesting possibility
for Run IIa of a SUSY signal which contains real Higgs bosons. 
The Run IIb 3$\sigma$ dicovery reaches quoted above for the 
$\gamma bb \met$ and $bbbb \met$ channels correspond to 
signal times branching ratio cross sections of 
1 fb and 4 fb respectively. 
For the parameters of Fig. 3 the maximum reach in Higgs
mass then corresponds to just over 145 GeV and 115 GeV respectively. 

In order to identify the Higgs boson directly in a sample
of events arising from Higgsino decays it is necessary to 
observe a peak in the $bb$ invariant mass. 
The identifiable di-boson final states and 
large $\met$ carried by the Goldstinos
render the supersymmetric Higgs boson final states discussed
here relatively clean.
Reconstructing the Higgs mass peak should be
relatively straightforward 
compared to SM $Wh$ and $Zh$ production modes
which suffer from much larger continuum $bb$ backgrounds. 

All the new signatures presented here involve hard photons, 
leptons, and/or $b$-jets, in association with significant missing energy. 
New triggers are therefore not required, but 
final state specific off line analysis
should be implemented in order to search for supersymmetry and/or
the Higgs boson in these interesting channels. 

Finally, Higgsino decay with a measurable macroscopic decay length
to the Goldstino would render all the di-boson
final states discussed here essentially background free. 
A search for such final states 
requires a special analysis for displaced 
$\ell^+ \ell^-$, $jj$, or $bb$ with large invariant mass 
and approximately uniform angular distribution with respect
to the beam axis \cite{mesino}.

\section*{Acknowledgements}
We would like to thank R.~Demina, G.~Landsberg, and J.~Wells for 
invaluable discussions. 
The work of K.~M.~was supported by Fermilab which 
is operated under DOE contract DE-AC02-76CH03000, 
and that of S.~T.~by 
the US National Science Foundation under grant PHY98-70115, 
the Alfred P. Sloan Foundation, and Stanford
University through the Frederick E.~Terman Fellowship.


\end{document}